# A FinTech Clustering Framework: Technology, Model, and Stakeholder Perspectives

Pak-Lok Poon [a, *], Santoso Wibowo [a], Sau-Fun Tang [a]

[a] *School of Engineering and Technology, Central Queensland University, 120 Spencer Street, Melbourne, VIC 3000, Australia*

**ABSTRACT**

Nowadays, the global booming of FinTech can be seen everywhere. FinTech has created innovative disruptions to traditional, long-established financial institutions (e.g., banks and insurance companies) in financial services markets. Despite of its popularity, there are many different definitions of FinTech. This problem occurs because many existing studies only focus on a particular aspect of FinTech without a comprehensive and in-depth analysis. This problem will hinder further development and industrial application of FinTech. In view of this problem, we perform a narrative review involving over 100 relevant studies or reports, with a view to developing a FinTech clustering framework for providing a more comprehensive and holistic view of FinTech. Furthermore, we use an Indian FinTech firm to illustrate how to apply our clustering framework for analysis.

---

\* Corresponding author. Email: p.poon@cqu.edu.au

## Introduction

Financial Technologies (FinTechs), such as SWIFT and Bloomberg, have been around for decades, but only over the last few years they have revolutionized the way people interact with financial services (Mention, 2021; Pousttchi & Dehnert, 2018; Puschmann, 2017). Although it is almost certain that technology advancement has an impact on the financial industry, what make the FinTech revolution so unique are: (a) the pace at which new technologies are tested and introduced into the financial industry is much faster than ever before, and (b) much of the change is happening from outside the financial industry, where young start-ups and large established technology companies are generally attempting to disrupt the incumbents by introducing new products and technologies (Chaklader et al., 2023; Goldstein et al., 2019).

From a review of literature on FinTech (Harrington, 2017; Schueffel, 2016; Thakor, 2020), we have observed a problem. When people (particularly those working in the industry) talk about FinTech today, they generally only have a limited understanding of the concept (Zavolokina et al., 2016). There is still a lack of consensus on the definition of FinTech among researchers and practitioners (Milian et al., 2019). A major reason for this incomplete and, perhaps, imprecise understanding of FinTech is that many existing studies only focus on a particular aspect of FinTech without a comprehensive and in-depth analysis. This problem is likely to hinder further development and industrial application of FinTech. With a view to addressing this problem, we develop a clustering framework for interconnected FinTech perspectives, through which a holistic overview of FinTech can be provided.

## Study Approach

Consideration has been made regarding whether the study should be a systematic review or a narrative review. These two types of review mainly differ in their objectives and methods (Collins & Fauser, 2005).

A *systematic review* formulates a well-defined research question, then uses qualitative and quantitative methods to analyze all the evidence for answering the research question. Thus, this review has a "narrow" focus of the research question. Also, a systematic review involves detailed and comprehensive literature searches with the use of a criterion-based selection of relevant evidence. Although systematic reviews are popular in the research community, Collins and Fauser (2005) argue that its "narrow" focus and prescribed methods do not allow for "comprehensive" coverage.

On the other hand, a *narrative review* is a scholarly summary along with interpretation and critique. This review is generally comprehensive and covers a wide range of issues within a given topic. A narrative review often only has a topic of interest without predetermined research questions. Also, this review does not necessarily state or follow rules about the search for evidence (i.e., without a specified search strategy), and does not involve prescribed databases for literature search.

*This study adopted the approach of a narrative review* because of two reasons: (a) a historical review was indispensable for tracing the development of FinTech, but the narrative thread could be lost in the strict rules of a systematic review; and (b) a broad examination of various aspects of FinTech was not possible due to the restrictive focus of a systematic review.

Despite using a narrative review without the need for a rigorous and structured literature search and prescribed databases (Jahan et al., 2016), our study still used an "informal" search strategy. A total of 25 FinTech-related "preliminary" articles were initially collected from two FinTech studies (Poon et al., 2024a; 2024b). We then followed the relevant references (e.g., other journal and conference papers) mentioned in these "preliminary" articles to find more relevant ones. To supplement this snowballing approach, we searched for more relevant information online, using search words such as "FinTech", "financial technology", "financial technologies", "financial innovation", and "disruptive technologies".

Because the current study also aimed at investigating the development of FinTech in the financial services industry and the practitioner's views on FinTech, the study covered both the academic literature as well as information from various industry sources (e.g., practitioner's journals/magazines and industry/government reports). While academic literature enables rigorous knowledge synthesis of the various aspects of FinTech, information from industry sources provides an overview of the FinTech industry and its environment. All in all, the current study involved 76 and 27 FinTech-related references from academic and practitioner sources, respectively.

## Different Definitions of FinTech

The continuous revolution of FinTech has resulted in multiple definitions of FinTech, with each definition emphasizes on some (but not all) aspects of FinTech. These definitions are grouped into three perspectives as follows:

### Technology Perspective

It focuses on *FinTech-enabled technologies* (or simply *FinTech technologies*). This perspective acknowledges that a FinTech innovation can be assessed in terms of technology by considering this innovation as the practical application of technical processes or methods (Gatignon et al., 2016).

- *Technology* used to provide financial markets a financial product or service, characterized by advanced technology relative to existing technology in that market (Knewtson & Rosenbaum, 2020).
- The use of *technology* to provide new and improved financial services (Gomber et al., 2018).
- A broad category that encompasses many different *technologies* for changing the way consumers and businesses access their finances and compete with traditional financial services (Peek, 2020).

### Model Perspective

It focuses on *FinTech models, ideas, innovations, applications,* and *businesses* (thereafter, they are collectively known as "FinTech models"). This perspective focuses on how FinTech companies take advantage of changing customer demands and expectations on financial services via their company's FinTech model innovations (Mastropietro, 2022).

- Any innovative *ideas* that improve financial services processes by employing technology solutions to different business situations, and these *ideas* could also lead to new business *models* or even new *businesses* (Leong & Sung, 2018).

- The *application* of *technological innovations* to financial services and processes (Lagna & Ravishankar, 2021).
- The new business *model* for the global financial sector, offering clear and enormous potential for vast economies of scale and scope, massive cost savings and efficiency gains, significant risk reduction, and opening the door to banking for countless currently unbanked people (Fischer, 2021).

### *Stakeholder Perspective*

Examples of FinTech *stakeholders* are FinTech start-ups, traditional financial institutions, technology developers, Tech Titans, government, and financial customers. This perspective acknowledges that FinTech stakeholders not only affect the survival and development of FinTech companies, but also determine the activities and effectiveness of FinTech innovations in these companies (Ya & Rui, 2006).

- Technology-enabled new *entrants* that change how financial services are structured provisioned, and consumed (McWaters & Galaski, 2017).
- *Organizations* combining innovative business models and technology to enable, enhance, and disrupt financial services (CEPS, 2023).
- An *ecosystem* of heterogenous, non-linear, dynamic, and complex network of agents that interact with each other to provide a wide range of financial products and services to end customers (Lagna & Ravishankar, 2021; Lee & Shin, 2018; Muthukannan et al., 2020; Senyo et al., 2022).

To alleviate the confusion caused by the various FinTech definitions used by different researchers and practitioners, the next section systematically discusses the important concepts of FinTech in terms of the three perspectives as shown in Fig. 1.

## A Clustering Framework for FinTech Perspectives

In view of all the FinTech definitions mentioned in the preceding section, we developed a FinTech clustering framework involving three interconnected perspectives as shown in Fig. 1. In this figure, we argue that technologies (the left block) serve as the underpinning enabler for the adoption and implementation of various FinTech models, ideas, innovations, applications, and businesses (the middle block). For example, Artificial Intelligence (AI) and Machine Learning (ML) are two main technologies that underpin the development of robo-advisors for supporting the wealth management model in FinTech. Then, FinTech models (the middle block) create new business opportunities to or impacts on different stakeholders (the right block). For example, the digital banking model in FinTech allows challenger banks to enter the financial services market by providing services to customers with banking fees lower than that of traditional, long-established banks.

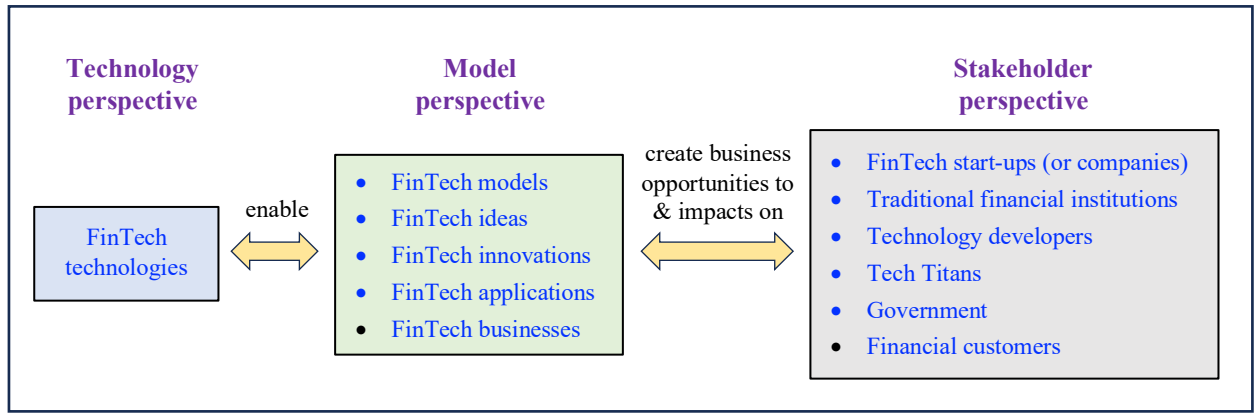

**Figure 1.** A clustering framework of FinTech perspectives.

## *Technology Perspective*

Table 1 lists 21 major FinTech technologies, among which AI/ML, big data/data (or predictive) analytics, and blockchain/cryptocurrency are frequently mentioned. On the other hand, the FinTech technologies that are relatively less mentioned are Application Programming Interface (API), Open-Source Software (OSS), quantum computing, Quick Response (QR) code, virtual card, voice technology, and 5G network.

Table 1 is developed based on our narrative review of the relevant literature. Thus, the number of occurrences of a particular FinTech technology in the references listed in Table 1 may not necessarily reflect the degree of popularity of this technology in FinTech. Nevertheless, Table 1 still provides a useful overview of a variety of FinTech technologies.

Bholat (2020) argues that the popularity of AI/ML in FinTech is caused by the variety, volume, and velocity of data created in the Internet era. Buchanan and Wright (2021) argue that the tremendous growth of using AI/ML in FinTech is attributed by their wide range of applications in fraud detection and compliance monitoring, credit scoring, financial distress prediction, robo-advising, and algorithmic trading. AI is predicted to contribute up to US$ 15.7 trillion to global GDP by 2030, with the financial services industry becoming an area for substantial activity (Rahman, 2020).

**Table 1.** Different FinTech technologies.

| FinTech technologies | References |
|---|---|
| Application programming interface (API) | Ünsal et al., 2020 |
| AI/ML | Cao et al., 2021; Mirza et al., 2023; Nguyen et al., 2022 |
| Augmented reality (AR)/virtual reality (VR) | Begum et al., 2022 |
| Big data/data (or predictive) analytics | Cao et al., 2021; Nguyen et al., 2022; Yin et al., 2022 |
| Biometrics | Wang, 2021 |
| Blockchain/cryptocurrency | Lee and Lim, 2021; Mohanta et al., 2018; Mosteanu and Faccia, 2021 |
| Cloud computing technology | Cheng et al., 2022; Meng et al., 2021 |
| Crowdfunding platform | Hoegen et al., 2018; Wonglimpiyarat, 2018 |
| Cybersecurity technologies | Mani, 2019; Selvaraj, 2021 |
| Internet of Things (IoT)/sensors | Maiti and Ghosh, 2023 |
| Near field communication (NFC) | Au and Kauffman, 2008; Halaweh and Al-Qaisi, 2016 |
| No-code (or low-code) development platform | Clarke, 2021; Clere, 2022 |
| Open-source software (OSS) | Ferreira, 2022 |
| Quantum computing | Mosteanu and Faccia, 2021 |
| Quick response (QR) code | Eren, 2024 |
| Robotic process automation (RPA) | Thekkethil et al., 2021; Villar and Khan, 2021 |
| Sentiment Analysis | Warjiyono et al., 2019; Widiantoro et al., 2021 |
| Smart contract | Lee, 2022; Lo et al., 2021 |
| Virtual card | Singh, 2023 |
| Voice technology | Chen et al., 2022 |
| 5G network | Ris, 2021 |

Khatri et al. (2021) argue that big data is useful for developing FinTech innovations, and inherently incentivizes, exposes, and resolves FinTech challenges. Haidar (2020) argues that big data is a key enabler of FinTech because it facilitates financial companies to innovate and improve their services and offerings to earn customer loyalty and surpass their competitors. For example, big data enables financial companies to perform the typically protracted and expensive credit-risk scoring and assessment tasks faster.

By their very nature, blockchain and cryptocurrency are closely related to payment transactions (Han et al., 2023), and thus have immediate contributions to the financial services industry. The popularity of these technologies in FinTech is also caused by two main reasons. First, blockchain can significantly shorten the settlement period and accelerate the payment process (Lewis et al., 2017). Second, using blockchain and cryptocurrency (e.g., bitcoin) in financial services can effectively prevent adverse behavior and repercussions, such as double spending and forgery (Barber et al., 2012; Nofer et al., 2017).

Nowadays, 5G network is commonly considered as the next generation of wireless data networks (Attaran, 2023). 5G is argued to be a gateway to the new era of financial industry. The contribution of 5G to FinTech is mainly due to its high transmission speed (up to 10 gigabits per second, which is about 100 times faster than 4G). In general, 5G contributes to FinTech in several aspects such as: (a) real-time mobile-banking user experience, (b) streamlined lending, (c) security and fraud detection, and (d) an enabler for AI, IoT, and VR (England, 2022).

### *Model Perspective*

This subsection discusses some popular FinTech models, their concepts, value propositions, underlying technologies, and associated applications.

*Payment Model:* As payments are daily financial activities, those financial services companies adopting this model (known as PayTechs) will be able to attract customers quickly. The main value created by this model is streamlined payments experienced by customers in terms of speed, convenience, and multi-channel accessibility. This model even makes payments to be embedded (and, hence, "invisible") within customer journeys, by reducing the obstacles of making a payment transaction to a minimum. This explains why PayTechs currently account for about 25% of FinTechs and they focus on the payment value chain, payment facilitation, and new payment propositions (Gancz et al., 2022).

BNY Mellon (2015) has identified several FinTech applications associated with this payment model: mobile wallet, P2P payment, foreign exchange and remittance, real-time payment, and *digital currency solution* (this solution offers banks a new way to handle, manage, and distribute their funds in the form of digital money, designed to beneficially replace fiat currencies through blockchain and process automation).

Various FinTech technologies are used to support this model including, for example, blockchain/cryptocurrency, cloud computing, cybersecurity technologies, NFC, QR code, payment gateways (e.g., PayPal, SecurePay, and Alipay), virtual card, and mobile card reader.

Another emerging technology that underpins the payment model is *biometric authentication*, where payments can be made using customers' fingerprint, facial and iris recognition, heartbeat analysis, and vein mapping.

*Digital Banking Model:* This model is almost identical to that of a traditional, long-established bank ("high street bank") with physical branches, except that with the huge cost savings in manpower and real estate. In this model, *challenger banks* (e.g., N26) offer no-frills individual and business bank accounts through a well-defined digital infrastructure (i.e., challenger banks offer services via an app or through their websites). Customers of challenger banks not only benefit from higher interest rates on their savings and reduced banking fees, but they can also enjoy additional services (e.g., real-time spending notifications, and personalized advice and analytics) that are not available in high street banks.

Another motivating factor to make customers move from a high street bank to a challenger bank is "personalization". Relatively speaking, challenger banks have higher agility and speed in innovation to meet customers' needs in terms of their lifestyle choices. Even if customers' needs constantly change, challenger banks will be able to offer more that is hyper-personalized and highly appealing (e.g., simple account opening and operation) (Dua, 2023).

GlobeNewswire (2020) projected that the market size of the challenger bank industry will reach $471.0 billion by 2027. Examples of the technologies that underpin digital banking are open banking API, AI, ML, blockchain, advanced data analytics, and mobile apps (Serheichuk, 2021).

*Wealth Management Model:* Traditionally, banks and wealth managers mainly offer expensive financial products with very high minimums. Thus, these products are mainly for high-net-worth individuals. Consequently, small investors are often excluded from these wealth management services. Nowadays, with the availability of wealth management applications (e.g., robo-advisors and digital brokerage), financial services are made available at a fraction of the cost of a human financial advisor (Jung et al., 2018).

Consider, for example, robo-advisors. Investors (especially the small ones) prefer robo-advisors more than human financial advisors, because these application tools are less vulnerable to potential conflicts of interest, have significant lower and more transparent cost structures, and are much less prone to misguided incentive-based compensation schemes and conflicting kickback-payments (Brenner & Meyll, 2020). The total value of assets managed by robo-advisors is expected to reach a staggering $4.6 trillion by 2022 (Meola, 2021). The two main technologies underpinning robo-advisors are AI-based algorithms and ML.

Apparently, robo-advisors and human financial advisors are competitors of each other. This view, however, has been changing. Nowadays, traditional banks view robo-advisors as "healthy" due to several reasons: (a) robo-advisors represent a challenge to the traditional banks so that the latter are motivated to offer better customer experiences and journey; and (b) combining the reach and tech-enabled capabilities of digital platforms, with the relationship and client base of traditional wealth managers, create synergies that allow a wider group of clients to be served more efficiently in terms of product access and cost (Ng, 2023).

*Crowdfunding Model:* This model provides an effective mechanism to fund a project or venture by raising many small amounts of money from a huge number of people, typically through the Internet (Hoegen et al., 2018). Crowdfunding is particularly advantageous to small businesses and start-ups by offering them an opportunity to succeed through demonstrating their innovate business models to the world.

Crowdfunding involves three parties: the project initiator who needs funding, the contributors who may be interested in supporting the project, and the moderating organization that facilitates the engagement and collaboration between the initiator and the contributors. There are three major crowdfunding models: rewards-based, donation-based, and equity-based (Lee & Shin, 2018). These three models are technically supported by the Internet-based crowdfunding platforms such as Kickstarter and Crowdfunder. Note that many FinTech *companies* have the same names as the *applications* or *platforms* they offer (e.g., Kickstarter). Despite this, readers should know what they are referring to from the context.

In general, crowdfunding is associated with several advantages and disadvantages (Verified Payments, 2022). On the positive side, crowdfunding offers three major benefits:

- Whereas traditional fundraising needs significant effort to persuade investors who are mainly interested in the return of investment, crowdfunding offers an opportunity to raise money for any project or idea that is innovative and thought-provoking and with the potential to generate significant profit in the future.

- Business control and management remains in the hands of the project initiator.

- The project initiator can reach a large audience by publishing particular posts, videos, ads, and information. Once the idea becomes visible, genuine feedback can be obtained from those who are interested in supporting this idea. This feedback then allows the project initiator to tweak the initial idea.

On the negative side, crowdfunding has the following major weaknesses:

- Crowdfunding may take long time to perform research for identifying what audience to target and what features to offer to stand out from the crowd.

- Most FinTech start-ups are unable to raise sufficient funds for company growth.

- Crowdfunding takes longer time to raise fund. Often, most crowdfunding initiatives take weeks or even months to complete.

*Cash Flow Underwriting/Lending:* Cash flow underwriting provides a new, more accurate approach to evaluate borrowers' creditworthiness. This new approach analyzes real-time financial data, beyond the limitations of traditional credit reports. These credit reports often focus on *past* credit behavior. Cash flow underwriting, however, takes into consideration *real-time* data from a borrower's bank account, including income, spending patterns, and financial obligations (GDS Link, 2024). Thus, cash flow underwriting facilitates more informed lending decisions. A major factor of the growth of cash flow underwriting is the introduction of open banking, which allows secure and consent-based financial data sharing between banks and third-party providers.

Cash flow lending is related to cash flow underwriting, but there is major difference between the two. For cash flow lending, the lender who provides the loan assumes all finance risk. On the other hand, for cash flow underwriting, the underwriter determines the value of that risk for the lender.

The major underlying technologies used to support this model are: API, AI/ML, big data and predictive analytics, and cybersecurity technologies. Cash flow underwriting/lending are commonly used in payday and cash advance loans, property management, mortgage and auto financing, and credit card applications (Antosz, 2024).

*P2P Lending Model:* P2P lending is the model of lending money to individuals or businesses through online investment platforms (also called *intermediary P2P platforms*) that match lenders with borrowers, without an official financial institution (e.g., a bank) participating as an intermediary in the deal (Wang et al., 2021). These intermediary P2P platforms offer identity verification, proprietary credit models, loan servicing, as well as legal and compliance services to their customers (i.e., borrowers and lenders). A major difference between a bank and an intermediary P2P company is that the latter does not involve in the lending itself. Thus, an intermediary P2P company does not need to meet the stringent capital requirement as a bank. Big data, data analytics, and ML are the main enabling technologies for this model, which creates value to customers by using alternative credit scoring, online data sources, data analytics to price risks, rapid lending processes, and lower operating costs (Lee & Shin, 2018). There is a major difference between the P2P lending model and the crowdfunding model. The P2P lending model primarily focuses on debt consolidation and credit-card refinancing, but the crowdfunding model targets at providing funding for projects.

Bavoso (2022) reported that a larger share of loans has been originated through P2P platforms instead of traditional banking channels in the last decade. Policy makers generally prefer this trend because it contributes to better risk diversification by moving risks away from systemic financial institutions. However, this benefit comes with cost. Because the nature and role of P2P platforms have remained loosely defined, it is difficult to identify relevant regulatory challenges emerging from P2P lending (Bavoso, 2022).

*Capital Market Model: Capital markets* (*CMs*) are financial markets that bring buyers and sellers together to trade stocks, bonds, and other financial assets. Prominent players in FinTech-driven CMs are investment banks, custodians, exchanges, clearinghouses, and CM-focused information service providers (Teschner et al., 2016). In this model, the major focus areas are automation, data analytics and intelligence, and customer satisfaction through safe and convenient access. FinTech-driven innovations have created tremendous impacts on many parts of the CM's value chain such as investment, foreign exchange (forex) trading, and risk management.

Consider, for example, forex trading. Often forex traders are trading on thin time margins using an "intraday" strategy, because small hour-on-hour fluctuations on currency exchange rates can make a big difference. With the support of forex mobile apps (e.g., NetDania Stock and Forex Trader), forex traders can obtain the latest market news that may influence the currency rate movements, and perform more "responsive" real-time trading (Stan, 2018). These mobile apps also generate quotes and charts, and provide forex traders with access to their trading accounts at their fingertips at all times. Examples of the technologies underpinning these forex mobile apps are AI, ML, 4G/5G networks, and cybersecurity technologies.

*Insurance Services Model:* Insurance has now become ripe for disruption by Insurance Technology (InsurTech — an extension of FinTech beyond the banking sector) in much the same way as banking has by FinTech (Alt et al., 2018; Sosa & Montes, 2022). InsurTech companies aim at providing a more direct relationship between the insurers and the customers via a combination of mobile apps, wearables[1], and claims processing tools. InsurTech companies create values to their customers by providing: (a) an enriched connectivity with AI solutions, (b) personalized product offerings, (c) an exceptional digital customer experience, and (d) streamlined processes (Stoeckli et al., 2018). Two popular InsurTech apps are Lemonade and Hippo, which are supported by various technologies such as AI, ML, blockchain, big data, data analytics, IoT, and mobile apps.

---

[1] *Wearables* (or *wearable technology*) is a category of electronic devices that can be worn as accessories, embedded in clothing, implanted in the user's body, or even tattooed on the skin. These devices are hands-free gadgets, powered by microprocessors and enhanced with the ability to send and receive data via the Internet.

As InsurTech is part of FinTech in the same way insurance is part of finance, consumer trust plays an important role in the success of InsurTech. Zarifis and Cheng (2022) performed an empirical study, and found that consumers bring with them some pre-existing beliefs (or trust) on AI and related technologies (e.g., chatbots or virtual assistants) that underpin InsurTech. Therefore, consumer trust on InsurTech does not only base on their direct experience with InsurTech, but is also influenced by their existing beliefs (or trust) on AI and related technologies.

*Platform-Based Model:* More recently, we have witnessed a further extension of FinTech (besides InsurTech) — BigTech (Beard, 2022; Cornelli et al., 2023). JPMorgan (2022) reported that BigTech had a combined market value of about $2.5 trillion in 2022. The appearance of BigTech, together with digitization and platformization of finance, make the era of so-called *FinTech 4.0* (Arner et al., 2022).

The stakeholders of BigTech are known as *Tech Titans*. Examples of these Tech Titans are Google, Apple, Facebook, Amazon, and Microsoft (GAFAM) in the U.S., and Baidu, Alibaba, and Tencent (BAT) in China. These Tech Titans initially started their FinTech activities in the payment area (corresponding to the payment model discussed earlier) for streamlining their core businesses (e.g., e-commerce). Later, they leveraged their platforms to expand into vast ecosystems covering other areas such as lending, investment, and insurance (Bethlendi & Szöcs, 2022). These Tech Titans have captured dominant market share, allowing them to capitalize on network effects and leverage their core offerings as multi-sided platforms for commerce and innovation (Jones & Ozcan, 2021). Successful Tech Titans use the platform-based model to capitalize on "winner-takes-all" dynamics and strong reinforcing feedback loops, allowing them become synonymous with entire industries (Jones & Ozcan, 2021).

The entry of Tech Titans into the financial market is driven by several reasons: (a) diversifying their revenue streams, (b) accessing new sources of data, (c) complementing and reinforcing their core commercial activities, and (d) increasing their customer base and loyalty (FSB, 2019). Some underlying technologies behind this platform-based model are API, AI, big data, and cloud computing.

*The Emerging Trends of FinTech Models:* The appearance of InsurTech and BigTech has indicated that FinTech is now extended from the banking sector to other sectors (e.g., Insurance). Since there are many players in the FinTech space, there is a fierce competition to acquire new customers and keep existing ones. Such competition has caused FinTech companies to move beyond addressing financial needs ("product-centric") by offering ancillary services such as accounting and coaching ("customer-centric").

Recently, we have also witnessed the occurrence of a particular niche of the FinTech industry, namely *decentralized finance* (*DeFi*). Basically, DeFi is an emerging technology that "reshapes" financial services based on secure distributed ledgers similar to those used by cryptocurrencies (Gramlich et al., 2023; Zetzsche et al., 2020). DeFi allows users to perform financial transactions or peer-to-peer digital exchanges (e.g., transfers, lending, savings, investing, and trading) without the presence of an intermediary entity (e.g., banks and brokerages), thereby eliminating the fees that an intermediary entity charges for using its services. Any individuals holding money in a secure digital wallet with an Internet connection can get access to DeFi applications. This allows DeFi applications to be accessible across conventional boundaries, markets, regions, and different layers of society.

### *Stakeholder Perspective*

This perspective corresponds to the FinTech ecosystem, in which new and old stakeholders combine to offer unique capabilities that complement one another and contribute to innovation (Lagna & Ravishankar, 2021; Lee & Shin, 2018; Muthukannan et al., 2020; Senyo et al., 2022). With the introduction of various complementary technologies, the complexity of FinTech ecosystems is rapidly increasing as new stakeholders are emerging and new connections are established.

There are six major interacting stakeholders in the FinTech ecosystem as follows:

- *FinTech start-ups* offer technology-mediated services in payments, digital banking, wealth management, crowdfunding, P2P lending, capital market, and insurance (corresponding to the FinTech models discussed in the preceding subsections "Payment Model" to "Insurance Services Model") to create value to financial customers. FinTech start-ups generally adopt a strategy of "unbundling" financial services, which serves as a major driver of growth in the FinTech sector.

- *Traditional financial institutions* (e.g., large commercial banks and insurance companies) are a major driving force in the FinTech ecosystem. After recognizing the disruptive power of FinTech, traditional financial institutions respond to protect their interests by reinventing their products, processes, and business models (Drummer et al., 2017). These institutions previously considered fast-growing FinTech start-ups as threats. Today, these institutions have shifted their focus and strategies from competing to collaborating with FinTech start-ups with various funding provisions (Lee & Shin, 2018).

- *Government* establishes a stable regulatory infrastructure for the financial services market. For example, government can offer licensing of financial services, relaxation of capital requirements, and tax incentives to boost the growth and development of FinTech start-ups.

- *Financial customers* use and benefit from various FinTech services. These customers are also the source of revenue generations for FinTech companies. When compared with large organizations, individual customers and small-to-medium sized firms are the predominant revenue source for FinTech companies (Lee & Shin, 2018).

- *Technology developers* invent and provide different kinds of disruptive technologies (e.g., those listed in Table 1) to enable FinTech companies to launch their innovative services quickly and effectively. In return, the FinTech industry is generating revenue for these technology developers.

- *Tech Titans* increasingly underpin our social, political, and economic worlds by providing the digital infrastructure on which we rely to live our lives (similar to Big Oil and Big Banks in the past) (Birch & Bronson, 2022). According to a report by the World

Economic Forum, Tech Titans are more disruptive to banks than FinTech start-ups (Browne, 2017).

## Case Study: FlexiLoans

This section illustrates how to use our FinTech clustering framework to systematically and holistically analyze an app (FlexiLoans) developed by a FinTech firm in India (called FlexiLoans.com).

In India, its MSME (Micro, Small, and Medium Enterprise) sector (which contributes to about 30% of the country's GDP) is largely underserved. For example, less than 10% of India's MSMEs have access to loans from "traditional" financial institutions, because most MSMEs do not have conventional credit histories required by these financial institutions. This creates a large capital gap. In view of this issue, FlexiLoans.com was founded in 2016 to bridge this gap, particularly in Tier-2 and Tier-3 cities where access to financial funding is limited (Mathias, 2024; Meghani, 2023).

FlexiLoans is a mobile app that provides quick business loans without collateral to India's MSMEs for supporting their working capital requirements. FlexiLoans offers unsecured business loans of various types (e.g., working capital loans, vendor financing, merchant cash advance, and line of credit) starting at a very low interest rate of 1% per month. By leveraging alternative data sources, the app offers a fast process of getting business loans approved within 48 hours, with no hidden charges and a hassle-free and paperless process. The app involves a simple three-step process: (a) download the app, (b) upload photos of some key documents, and (c) connect to the applicant's bank account.

### Technology Perspective

There are several FinTech technologies that underpin FlexiLoans. Examples of the technologies include the following:

- *API:* This technology enables FlexiLoans to integrate its lending platform with the Amazon Lending Marketplace in India. This integration allows: (a) Amazon sellers to apply and monitor their loans with FlexiLoans.com from their Amazon seller dashboard (Agarwal, 2019), and (b) FlexiLoans.com keeps its cost of acquisition low because the Amazon platform already has a huge amount of data about sellers (e.g., monthly sales and growth in sales).

- *Data-driven AI/ML technologies:* The in-house developed AI/ML technologies are able to read and process a large volume of images of uploaded documents in seconds. Besides image processing, these technologies can also solve complex problems such as credit scoring, creditworthiness analysis, and financial analysis.

- *Data analytics:* It empowers end-to-end risk assessments and facilitates real-time credit decisions. Coupled with the in-house developed AI/ML technologies, effective credit models and customer scorecards can be built which access diverse third-party data for comprehensive evaluation, reducing friction, enhancing processing speed, and improving customer experience.

- *Cybersecurity technologies:* All applicants' data are transferred over a secure connection to avoid unauthorized data disclosure and tampering during transmission.

### Model Perspective

FlexiLoans.com has co-created underwriting models with its co-lenders. Thus, the firm is adopting the "Cash Flow Underwriting/Lending" model. The firm also adopts the "Platform-Based" model. Through the firm's digital platform, merchants have access to a massive pool of capitals, from FlexiLoans.com as well as its co-lenders (including banks and other Non-Banking Financial Companies (NBFCs)). Today, FlexiLoans.com has over 150 ecosystem partners, including online retailers of fashion and beauty products, food-tech, pharm-tech, MSME SaaS (Software as a Service) platforms, point-of-sale firms, and other e-commerce giants (Meghani, 2023).

### Stakeholder Perspective

Various types of stakeholders are involved in the business of FlexiLoans.com as follows:

- *Traditional financial institutions:* FlexiLoans.com is supported by an advisory team of senior Risk and Credit professionals from leading banks and NBFCs. To support its operations, the firm has raised debts from financial institutions such as JM Financial and Vivriti Asset Management. Also, FlexiLoans.com is backed by a NBFC called Epimoney.
- *Government:* As an online lending FinTech platform in India, FlexiLoans.com is subject to the regulation and monitoring of the India's Government.
- *Financial customers:* As stated earlier, major customers of FlexiLoans.com are MSMEs in India. As of today, the firm has disbursed over 75,000 loans, with a total amount of about 5,000 crores Indian Rupee.
- *Technology developers:* In the past few years, it was observed that over 80% of borrowers via FlexiLoans have been using a mobile device to apply for a loan. Thus, FlexiLoans.com has formed a partnership with True Software Scandinavia AB (a privately held company headquartered in Sweden which has developed the Truecaller app). Truecaller is a smartphone app that helps improve user experience on all platforms, especially mobile. With Truecaller, user experience of FlexiLoans is improved with a one-click registration and handy identity verification (Exchange4media, 2019).
- *Tech Titans:* FlexiLoans.com has partnered with major e-commerce and payments platforms such as Amazon, Flipkart, Paytm, and PhonePe, enabling seamless loan origination within these ecosystems.

The above discussion has demonstrated the effectiveness and practicality of our FinTech clustering framework for analyzing the various aspects associated with a FinTech firm (including its ecosystem).

## Conclusion

To alleviate the confusion caused by many different FinTech definitions from various studies, we developed a FinTech clustering framework with three interconnected perspectives: technology, model, and stakeholder. The Technology perspective focuses on various FinTech-enabled technologies. The Model perspective focuses on FinTech models, ideas, innovations, applications, and businesses. The Stakeholder perspective includes various FinTech stakeholders such as FinTech start-ups, traditional financial institutions, technology developers, Tech Titans, government, and financial customers. These three perspectives are not isolated. FinTech technologies serve as the underpinning enabler for the adoption and implementation of various FinTech models, and in turn these FinTech models create new business opportunities to or impacts on different stakeholders. We also illustrated how to use our clustering framework for analysis using an Indian FinTech firm.

Not only our FinTech clustering framework provides a more comprehensive and holistic view of FinTech, but it also provides two practical implications to FinTech practitioners. First, it indicates that FinTech entrepreneurs and practitioners must have a thorough understanding of various contemporary FinTech technologies, before they can recognize the business opportunities brought forward by these technologies (Kreuzer et al., 2022). These business opportunities then lead to the subsequent formation of the corresponding FinTech models, through which sustainability in business success can be achieved. Second, analyzing FinTech models cannot be performed in vacuum. Rather, such analysis should be performed with respect to the relevant stakeholders, because it is these stakeholders who largely determine the success of a FinTech model (Gray & Purdy, 2018).

# References

Agarwal, P. (2019). *FlexiLoans Extends Partnership with Amazon Lending Marketplace to Provide MSME Credit.* Retrieved from https://yourstory.com/smbstory/flexiloans-partnership-amazon-msme-lending

Alt, R., Beck, R., & Smits, M. T. (2018). FinTech & the transformation of the financial industry. *Electronic Markets*, *28*, 235–243.

Antosz, D. (2024). *Cash Flow Underwriting: 5 Ways Lenders Can Drive Growth.* Retrieved from https://plaid.com/resources/lending/cash-flow-underwriting/

Arner, D. W., Buckley, R., Charamba, K., Sergeev, A., & Zetzsche, D. (2022). Governing FinTech 4.0: BigTech, platform finance, & sustainable development. *Fordham Journal of Corporate & Financial Law*, *27*(1), 1.

Attaran, M. (2023). The impact of 5G on the evolution of intelligent automation & industry digitization. *Journal of Ambient Intelligence & Humanized Computing*, *14*, 5977–5993.

Au, Y. A., & Kauffman, R. J. (2008). The economics of mobile payments: Understanding stakeholder issues for an emerging financial technology application. *Electronic Commerce Research & Applications*, *7*(2), 141–164.

Bank of New York (BNY) Mellon (2015). *Innovation in Payments: The Future is Fintech.* Retrieved from http://www.spainfinancialcentre.com/sites/default/files/innovation-in-payments_the-future-is-fintech._bny_mellon.pdf

Barber, S., Boyen, X., Shi, E., & Uzun, E. (2012). Bitter to better — How to make bitcoin a better currency. In A. D., Keromytis (Ed.), *Financial cryptography & data security* (pp. 399–414). Heidelberg: Springer.

Bavoso, V. (2022). Financial intermediation in the age of FinTech: P2P lending & the reinvention of banking. *Oxford Journal of Legal Studies*, *42*(1), 48–75.

Beard, A. (2022). Can Big Tech be disrupted? *Harvard Business Review*. Retrieved from https://hbr.org/2022/01/can-big-tech-be-disrupted

Begum, A., Rahaman, S., & Gaytan, J. C. T. (2022). Fintech & leverage of virtual reality: An outlook from UAE. In *Proceedings of the International Conference on Cyber Resilience.* Dubai, United Arab Emirates.

Bethlendi, A., & Szöcs, Á. (2022). How the Fintech ecosystem changes with the entry of Big Tech companies. *Investment Management & Financial Innovations*, *19*(3), 38–48.

Bholat, D. (2020). The impact of machine learning & AI on the UK economy — Conference overview. *Conference on the Impact of Machine Learning & AI on the UK Economy.* London, U.K.

Birch, K., & Bronson, K. (2022). Big Tech. *Science as Culture*, *31*(1), 1–14.

Board of Innovation (n.d.). *10 Innovative FinTech Business Models.* Retrieved from https://www.boardof innovation.com/blog/10-innovative-fintech-business-models/

Brenner, L., & Meyll, T. (2020). Robo-advisors: A substitute for human financial advice? *Journal of Behavioral & Experimental Finance*, *25*, 100275.

Browne, R. (2017). *Tech Giants Like Amazon & Facebook More Disruptive to Banks Than Fintech Start-ups: WEF.* Retrieved from https://www.cnbc.com/2017/08/22/tech-giants-more-disruptive-to-banks-than-fintech-startups.html

Buchanan, B. G., & Wright, D. (2021). The impact of machine learning on UK financial services. *Oxford Review of Economic Policy*, *37*(3), 537–563.

Cao, L., Yang, Q., & Yu, P. S. (2021). Data science & AI in FinTech: An overview. *International Journal of Data Science & Analytics*, *12*, 81–99.

Centre for European Policy Studies (CEPS) (2023). *Establishment of FinTech Supportive Environment.* Retrieved from https://www.ceps.eu/ceps-projects/establishment-of-fintech-supportive-environment/

Chaklader, B., Gupta, B. B., & Panigrahi, P. K. (2023). Analyzing the progress of FINTECH-companies & their integration with the new technologies for innovation & entrepreneurship. *Journal of Business Research*, *161*, 113847.

Chen, H., Chen, S., & Zhao, J. (2022). Integrated design of financial self-service terminal based on artificial intelligence voice interaction. *Frontiers in Psychology*, *13*, 850092.

Cheng, M., Qu, Y., Jiang, C., & Zhao, C. (2022). Is cloud computing the digital solution to the future of banking? *Journal of Financial Stability*, *63*, 101073.

Clarke, G. (2021). *Veritran Explain How Low-Code Technology Helps Banking Across LatAm.* Retrieved from https://thefintechtimes.com/veritran-explains-how-low-code-technology-helps-banking-across-latam/

Clere, A. (2022). *Low-Code Fintech Platform Toqio Raises €20mn in Series A.* Retrieved from https://fintechmagazine.com/articles/low-code-fintech-platform-toqio-raises-20mn-in-series-a

Collins, J. A., & Fauser, B. C. J. M. (2005). Balancing the strengths of systematic & narrative reviews. *Human Reproductive Update*, *11*(2), 103–104.

Cornelli, G., Frost, J., Gambacorta, L., Rau, P. R., Wardrop, R., & Ziegler, T. (2023). Fintech & big tech credit: Drivers of the growth of digital lending. *Journal of Banking & Finance*, *148*, 106742.

Drummer, D., Feuerriegel, S., & Neumann, D. (2017). Crossing the next frontier: The role of ICT in driving the financialization of credit. *Journal of Information Technology*, *32*(3), 218–233.

Dua, A. (2023). *Digital Banking: Traditional Banks Versus Fintech Providers.* Retrieved from https://fintechmagazine.com/articles/digital-banking-traditional-banks-versus-fintech-providers

England, J. (2022). *Fintech, IoT & 5G: How Connectivity is Transforming Fintech.* Retrieved from https://fintechmagazine.com/financial-services-finserv/fintech-iot-5g-how-connectivity-is-transforming-fintech

Eren, B. A. (2024). QR code m-payment from a customer experience perspective. *Journal of Financial Services Marketing*, *29*, 106–121.

Exchange4media (2019). *Flexiloans partners with Truecaller to provide fast track user on-boarding.* Retrieved from https://www.exchange4media.com/marketing-news/flexiloans-partners-with-truecaller-to-provide-fast-track-user-on-boarding-96621.html

Ferreira, I. (2022). *Fintech & Open Source.* Retrieved from https://www.fintechna.com/articles/fintech-and-open-source/

Financial Stability Board (FSB) (2019). *BigTech in Finance: Market Developments & Potential Financial Stability Implications.* Retrieved from https://www.fsb.org/wp-content/uploads/P091219-1.pdf

Fischer, M. (2021). *Fintech business models: Applied Canvas method & analysis of venture capital rounds.* Berlin: De Gruyter.

Gancz, A., Halder, D., Partelow, P., & Elinson, S. (2022). *How the Rise of PayTech is Reshaping the Payments Landscape.* Retrieved from https://www.ey.com/en_au/payments/how-the-rise-of-paytech-is-reshaping-the-payments-landscape

Gatignon, H., Gotteland, D., & Haon, C. (2016). Assessing innovations from the technology perspective. In H., Gatignon, D., Gotteland, & C. Haon (Eds.), *Making innovation last* (pp. 19–51), vol. 1. London: Palgrave Macmillan.

GDS Link (2024). *The Rise & Challenges of Cash Flow Underwriting in FinTech.* Retrieved from https://www.gdslink.com/the-rise-and-challenges-of-cash-flow-underwriting-in-fintech/

GlobeNewswire (2020). *Global Neo & Challenger Bank Market (2020 to 2027) — By Service Type & End-User.* Retrieved from https://www.globenewswire.com/news-release/2020/09/10/2091869/0/en/Global-Neo-and-Challenger-Bank-Market-2020-to-2027-by-Service-Type-and-End-user.html

Goldstein, I., Jiang, W., & Karolyi, G. A. (2019). To FinTech & beyond. *Review of Financial Studies*, *32*(5), 1647–1661.

Gomber, P., Kauffman, R. J., Parker, C., & Weber, B. W. (2018). On the FinTech revolution: Interpreting the forces of innovation, disruption, & transformation in financial services. *Journal of Management Information Systems*, *35*(1), 220–265.

Gramlich, V., Guggenberger, T., Principato, M., Schellinger, B., & Urbach, N. (2023). A multivocal literature review of decentralized finance: Current knowledge & future research avenues. *Electronic Markets*, *33*, 11.

Gray, B., & Purdy, J. (2018). *Collaborating for our future: Multistakeholder partnerships for solving complex problems.* Oxford: Oxford University Press.

Haidar, B. (2020). *The Role of Big Data in the 2020's Fintech Revolution.* Retrieved from https://fintechmagazine.com/venture-capital/role-big-data-2020s-fintech-revolution

Halaweh, M., & Al-Qaisi, H. (2016). Adoption of Near Field Communication (NFC) for mobile payments in the UAE: A merchant's perspective. *International Journal of e-Business Research*, *12*(4), 38−56.

Han, H., Shiwakoti, R. K., Jarvis, R., Mordi, C., & Botchie, D. (2023). Accounting & auditing with blockchain technology & artificial intelligence: A literature review. *International Journal of Accounting Information Systems*, *48*, 100598.

Harrington, C. (2017). *Why Fintech Could Lead to More Financial Crime.* Retrieved from https://www.cfainstitute.org/en/research/cfa-magazine/2017/why-fintech-could-lead-to-more-financial-crime

Hoegen, A., Steininger, D. M., & Veit, D. (2018). How do investors decide? An interdisciplinary review of decision-making in crowdfunding. *Electronic Markets*, *28*, 339−365.

Jahan, N., Naveed, S., Zeshan, M., & Tahir, M. A. (2016). How to conduct a systematic review: A narrative literature review. *Cureus*, *8*(11), 864.

Jones, R., & Ozcan, P. (2021). *Rise of BigTech Platforms in Banking.* Retrieved from https://www.sbs.ox.ac.uk/sites/default/files/2021-02/Rise%20of%20BigTech%20 Platforms%20in%20Banking%20-%20Oxford%20White%20Paper%20Final%20% 28002%29.pdf

JPMorgan (2022). *The Future of Big Tech.* Retrieved from https://www.jpmorgan.com/ insights/research/future-of-big-tech

Jung, D., Dorner, V., Weinhardt, C., & Pusmaz, H. (2018). Designing a robo-advisor for risk-averse, low-budget consumers. *Electronic Markets*, *28*, 367−380.

Khatri, A., Singh, N. P., & Gupta, N. (2021). Big data analytics: Direction & impact on financial technology. Journal of Management, *Marketing & Logistics*, *3*, 218−234.

Knewtson, H., & Rosenbaum, Z. (2020). Toward understanding FinTech & its industry. *Managerial Finance*, *46*(8), 1043−1060.

Kreuzer, T., Lindenthal, A. K., Oberländer, A. M., & Röglinger, M. (2022). The effects of digital technology on opportunity recognition. *Business & Information Systems Engineering*, *64*, 47−67.

Lagna, A., & Ravishankar, M. N. (2021). Making the world a better place with fintech research. *Information Systems Journal*, *32*(1), 61102.

Lee, E. (2022). Technology-driven solutions to banks' de-risking practices in Hong Kong: FinTech & blockchain-based smart contracts for financial inclusion. *Common Law World Review*, *51*(1−2).

Lee, D. K. C., & Lim, C. S. L. (2021). Blockchain use cases for inclusive FinTech: scalability, privacy, & trust distribution. *The Journal of FinTech*, *1*(1), 2050003.

Lee, I., & Shin, Y. J. (2018). Fintech: Ecosystem, business models, investment decisions, & challenges. *Business Horizons*, *61*(1), 35−46.

Leong, K., & Sung, A. (2018). FinTech (Financial Technology): What is it & how to use technologies to create business value in Fintech way? *International Journal of Innovation, Management & Technology*, *9*(2), 74−78.

Lewis, R., McPartland, J., & Ranjan, R. (2017). Blockchain & financial market innovation. *Economic Perspectives*, *41*(7), 2−12.

Lo, S. W., Wang, Y., & Chuen, D. L. K. (2021). *Blockchain & smart contracts: Design thinking & programming for FinTech.* Singapore: World Scientific.

Maiti, M., & Ghosh, U. (2023). Next-generation Internet of Things in Fintech ecosystem. *IEEE Internet of Things Journal*, *10*(3), 2104–2111.

Mani, W. (2019). Cybersecurity & fintech at a crossroads. *ISACA Journal*, *2*, 1–7.

Mastropietro, B. (2022). *Fintech Business Models Overview.* Retrieved from https://www.coinspeaker.com/guides/fintech-business-models-overview/

Mathias, G. (2024). *FlexiLoans Secures $34.5m in Series C Led by Accion, Nuveen & Fundamentum.* Retrieved from https://ibsintelligence.com/ibsi-news/flexiloans-secures-34-5m-in-series-c-led-by-accion-nuveen-fundamentum/

McWaters, R. J., & Galaski, R. (2017). *Beyond Fintech: A Pragmatic Assessment of Disruptive Potential in Financial Services.* Retrieved from https://www3.weforum.org/docs/Beyond_Fintech_-_A_Pragmatic_Assessment_of_Disruptive_Potential_in_Financial_Services.pdf

Meghani, V. (2023). *How FlexiLoans Grew Quickly Yet Profitably.* Retrieved from https://www.forbesindia.com/article/take-one-big-story-of-the-day/how-flexiloans-grew-quickly-yet-profitably/88177/1

Meng, S., He, X., & Tian, X. (2021). Research on Fintech development issues based on embedded cloud computing & big data analysis. *Microprocessors & Microsystems*, *83*, 103977.

Mention, A.-L. (2021). The age of FinTech: Implications for research, policy & practice. *The Journal of FinTech*, *1*(1), 2050002.

Meola, A. (2021). *Top Robo Advisors in 2021: Performance Reviews, Returns, & Comparisons.* Retrieved from https://www.businessinsider.com/best-robo-advisors?IR=T

Milian, E. Z., de M Spinola, M., de Carvalho, M. M. (2019). Fintechs: A literature review & research agenda. *Electronic Commerce Research & Applications*, *34*, 100833.

Mirza, N., Elhoseny, M., Umar, M., & Metawa, N. (2023). Safeguarding FinTech innovations with machine learning: Comparative assessment of various approaches. *Research in International Business & Finance*, *66*, 102009.

Mohanta, B. K., Panda, S. S., & Jena, D. (2018). An overview of smart contract & use cases in blockchain technology. In *Proceedings of 9th International Conference on Computing, Communication & Networking Technologies.* Bengaluru, India.

Mosteanu, N. R., & Faccia, A. (2021). Fintech frontiers in quantum computing, fractals, & blockchain distributed ledger: paradigm shifts & open innovation. *Journal of Open Innovation: Technology, Market, & Complexity*, *7*(1), 19.

Muthukannan, P., Tan, B., Gozman, D., & Johnson, L. (2020). The emergence of a Fintech ecosystem: A case study of the Vizag Fintech Valley in India. *Information & Management*, *57*(8), 103385.

Ng, M. (2023). *Should Wealth Managers Fear Robo-Advisory?* Retrieved from https://fundselectorasia.com/should-wealth-managers-fear-robo-advisory/

Nguyen, D. K., Sermpinis, G., & Stasinakis, C. (2022). Big data, artificial intelligence & machine learning: A transformative symbiosis in favour of financial technology. *European Financial Management*, *29*(2), 517–548.

Nofer, M., Gomber, P., Hinz, O., & Schiereck, D. (2017). Blockchain. *Business & Information Systems Engineering*, *59*, 183–187.

Peek, S. (2020). *What is Fintech? Definition, Evolution & Examples.* Retrieved from https://www.uschamber.com/co/run/business-financing/what-is-fintech

Poon, P.-L., Pond, N. Y. L., & Tang, S.-F. (2024a). Employment & career prospects of technical-oriented jobs in the FinTech market. *Journal of Information Systems Education*, *35*(2), 203–217.

Poon, P.-L., Wibowo, S., Grandhi, S., & Tang, S.-F. (2024b). Investigating FinTech education & training in Australian universities. *Information Systems Education Journal*, *22*(1), 30–40.

Pousttchi, K., & Dehnert, M. (2018). Exploring the digitalization impact on consumer decision-making in retail banking. *Electronic Markets*, *28*, 265–286.

Puschmann, T. (2017). Fintech. *Business & Information Systems Engineering*, *59*(1), 69–76.

Rahman, M.A. (2020). *Artificial Intelligence Will Contribute Up to $15.7 Trillion to Global GDP by 2030.* Retrieved from https://medium.com/born-to-lead/artificial-intelligence-will-contribute-up-to-15-7-trillion-to-global-gdp-by-2030-3a51dbd90416

Ris, K. (2021). *5G & next-gen consumer banking services.* Boca Raton, FL: CRC Press.

Schueffel, P. (2016). Taming the beast: A scientific definition of Fintech. *Journal of Innovation Management*, *4*(4), 32–54.

Selvaraj, N. A. P. (2021). The essence of cybersecurity through Fintech 3.5 in preventing & detecting financial fraud: A literature review. *Electronic Journal of Business & Management*, *6*(2), 18–29.

Senyo, P. K., Karanasios, S., Gozman, D., & Baba, M. (2022). FinTech ecosystem practices shaping financial inclusion: The case of mobile money in Ghana. *European Journal of Information Systems*, *31*, 112–127.

Serheichuk, N. (2021). *Challengers & Neobanks: The Rise of Finance Alternatives.* Retrieved from https://www.n-ix.com/challenger-banks-neobanks/

Singh, G. (2023). *Payments: From Cashless to Cardless.* Retrieved from https://www.fintechweekly.com/magazine/articles/payments-from-cashless-to-cardless

Sosa, I., & Montes, Ó. (2022). Understanding the InsurTech dynamics in the transformation of the insurance sector. *Risk Management & Insurance Review*, *25*(1), 35–68.

Stan, A.-L. (2018). Computational speed & high-frequency trading profitability: An ecological perspective. *Electronic Markets*, *28*, 381–395.

Stoeckli, E., Dremel, C., & Uebernickel, F. (2018). Exploring characteristics & transformational capabilities of InsurTech innovations to understand insurance value creation in a digital world. *Electronic Markets*, *28*, 287–305.

Teschner, C., Bertali, V., Lavrov, B., Mikroulis, K., Rhode, W., Saumya, S., Vialaron, F., & Morel, P. (2016). *Fintech in Capital Markets: A Land of Opportunity.* Retrieved from https://www.bcg.com/publications/2016/financial-institutions-technology-digital-fintech-capital-markets

Thakor, A. V. (2020). Fintech & banking: what do we know? *Journal of Financial Intermediation*, *41*, 100833.

Thekkethil, M. S., Shukla, V. K., Beena, F., & Chopra, A. (2021). Robotic process automation in banking & finance sector for loan processing & fraud detection. In *Proceedings of the 9th International Conference on Reliability, Infocom Technologies & Optimization.* Noida, India.

Turguttopbas, N., & Kayral, I. E. (2023). Global peer-to-peer lending market. *PressAcademia Procedia*, *16*, 10–15.

Ünsal, E., Öztekin, B., Çavuş, M., & Özdemir, S. (2020). Building a fintech ecosystem: design & development of a fintech API gateway. In *Proceedings of the 2020 International Symposium on Networks, Computers & Communications.* Montreal, Canada.

Verified Payments (2022). *Is Crowdfunding The Best Way to Raise Funds for a Fintech Business?* Retrieved from https://verifiedpayments.com/blog/is-crowdfunding-the-best-way-to-raise-funds-for-a-fintech-business/

Villar, A. S., & Khan, N. (2021). Robotic process automation in banking industry: A case study on Deutsche Bank. *Journal of Banking & Financial Technology*, *5*(5), 71–86.

Wang, J. S. (2021). Exploring biometric identification in FinTech applications based on the modified TAM. *Financial Innovation*, *7*, 42.

Wang, Q., Xiong, X., & Zheng, Z. (2021). Platform characteristics & online peer-to-peer lending: Evidence from China. *Finance Research Letters*, *38*, 101511.

Warjiyono, Aji, S., Fandhilah, Hidayatun, N., Faqih, H., Liesnaningsih (2019). The sentiment analysis of Fintech users using support vector machine & particle swarm optimization method. In *Proceedings of the 7th International Conference on Cyber & IT Service Management.* Jakarta, Indonesia.

Widiantoro, A. D., Wibowo, A., Harnadi, B. (2021). User sentiment analysis in the Fintech OVO review based on the lexicon method. In *Proceedings of the 6th International Conference on Informatics & Computing.* Jakarta, Indonesia.

Wonglimpiyarat, J. (2018). Challenges & dynamics of FinTech crowd funding: An innovation system approach. *The Journal of High Technology Management Research*, *29*(1), 98–108.

Ya, S., & Rui, T. (2006). The influence of stakeholders on technology innovation: A case study from China. In *Proceedings of the 2006 IEEE International Conference on Management of Innovation & Technology* (pp. 295–299). Singapore.

Yin, F., Jiao, X., Zhou, J., Yin, X., Ibeke, E., Iwendi, M. G., & Biamba, C. (2022). Fintech application on banking stability using big data of an emerging economy. *Journal of Cloud Computing*, *11*, 43.

Zarifis, A., & Cheng, X. (2022). A model of trust in Fintech & trust in Insurtech: How artificial intelligence & the context influence it. *Journal of Behavioral & Experimental Finance*, *36*, 100739.

Zavolokina, L., Dolata, M., & Schwabe, G. (2016). FinTech — What's in a name? In *Proceedings of the 37th International Conference on Information Systems.* Dublin, Ireland.

Zetzsche, D. A., Arner, D. W., & Buckley, R. P. (2020). Decentralized finance. *Journal of Financial Regulation*, *6*(2), 172–203.